\documentclass[runningheads]{llncs}

\usepackage[T1]{fontenc}
\usepackage{graphicx}
\usepackage{todonotes}
\usepackage{xcolor}
\usepackage{listings}
\usepackage{jolie}
\usepackage{lstcoq}
\usepackage{hyperref}
\usepackage{cleveref}
\usepackage{placeins}

\pdfoutput=1

\newlang{coq}{Coq}{syellow}
\newcommand{\cc}{\lstinline[language=Coq,style=solarized-light]}
\newlang{haskell}{Haskell}{sgreen}
\newcommand{\hc}{\lstinline[language=Haskell,style=solarized-light]}
\newlang{jolie}{Jolie}{smagenta}
\newcommand{\jc}{\lstinline[language=Jolie,style=solarized-light]}

\newcommand{\dollar}{\mbox{\textdollar}}
\newcommand{\proc}[1]{\texttt{#1}}
\newcommand{\hacc}{\texttt{hacc}}
\newcommand{\eoe}{\hspace*\fill\ensuremath{\triangleleft}}

\begin{document}


\title{Certified Compilation of Choreographies with \hacc{}}


\author{
  Luís Cruz-Filipe \orcidID{0000-0002-7866-7484} \and
  Lovro Lugović \orcidID{0000-0001-9684-9567} \and
  Fabrizio Montesi \orcidID{0000-0003-4666-901X}
}

\authorrunning{L. Cruz-Filipe et al.}

\institute{
  Department of Mathematics and Computer Science, University of Southern Denmark
  \email{\{lcf, lugovic, fmontesi\}@imada.sdu.dk}
}

\maketitle

\begin{abstract}
  Programming communicating processes is challenging, because it requires writing separate programs that perform compatible send and receive actions at the right time during execution.
  Leaving this task to the programmer can easily lead to bugs.
  \emph{Choreographic programming} addresses this challenge by equipping developers with high-level abstractions for codifying the desired communication structures from a global viewpoint.
  Given a choreography, implementations of the involved processes can be automatically generated by \emph{endpoint projection (EPP)}.

  While choreographic programming prevents manual mistakes in the implementation of communications, the correctness of a choreographic programming framework crucially hinges on the correctness of its complex compiler, which has motivated formalisation of theories of choreographic programming in theorem provers.
  In this paper, we build upon one of these formalisations to construct a toolchain that produces executable code from a choreography.

\end{abstract}

\section{Introduction}

In traditional distributed programming, the programmer is tasked with writing the implementation of each \emph{process} (endpoint) as a separate program, taking care of correctly matching send and receive actions in the different programs.
This approach is known to be cumbersome and error-prone \cite{LLLG16}.

In \emph{Choreographic programming} \cite{M13p}, developers specify the desired communications between processes from a global viewpoint.
Given a choreographic program (called \emph{choreography}), correct implementations for all involved processes can be automatically generated by a procedure known as \emph{endpoint projection (EPP)}~\cite{M23}.
This avoids manual mistakes in the programming of communication actions, and provides important theoretical advantages, like deadlock-freedom by design---distributed code generated from a choreography is always deadlock-free, as choreographic languages do not have syntax for unmatched communications~\cite{CM13}.
In addition to these correctness advantages, this also saves time and lets the programmer focus on the bigger picture of the protocol being developed.

Defining and implementing EPP is technically involved~\cite{SY19}, which motivated mechanising theories of choreographic programming using interactive theorem provers~\cite{CMP21b,CMP21,HG22,PGSN22}.
Among these, the formalisation of \emph{Core Choreographies (CC)} and its EPP to the process calculus of \emph{Stateful Processes (SP)} was designed with broad applicability in mind~\cite{CMP22}.
Specifically, its design allows for annotating communications with arbitrary metadata and is parametric on the languages used to express local computation, data, process identifiers, etc.

In this work, for the first time, we reap the benefits of~\cite{CMP22} to develop \hacc{} (pronounced ``hack''): a tool for compiling choreographies in CC to executable code.
As target language for this executable code, we use the service-oriented programming language Jolie~\cite{MGZ14}.
Additionally, \hacc{} is designed with future extensibility in mind, so that new target languages can be added.

\begin{figure}[!b]
  \centering
  \includegraphics[width=.9\textwidth]{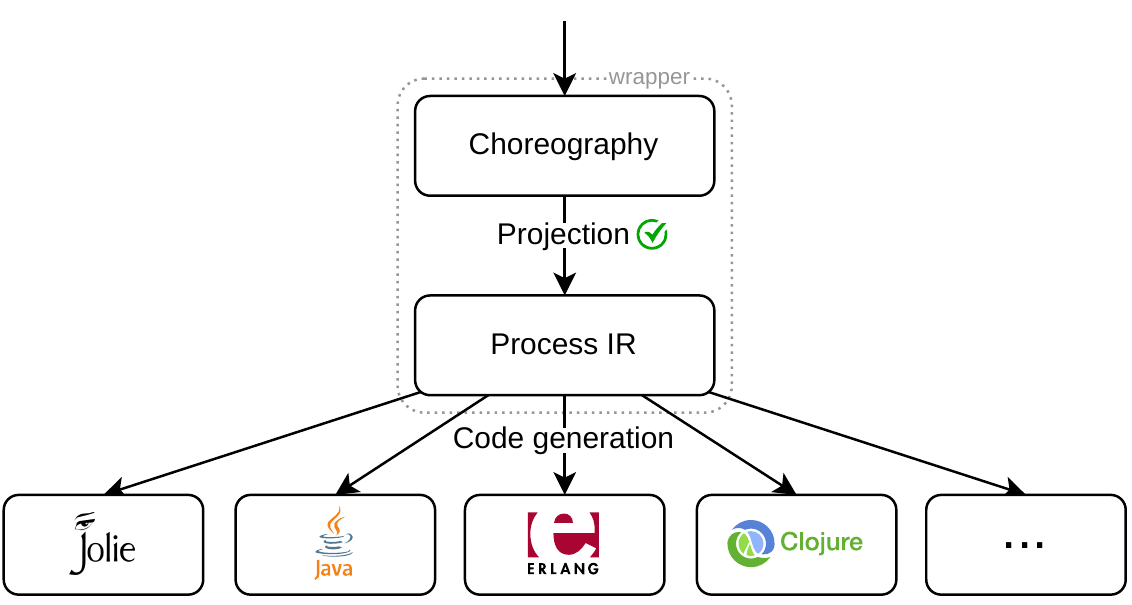}
  \caption{The architecture of \hacc{}'s compilation pipeline.}
  \label{fig:arch}
\end{figure}

The architecture of \hacc{} consists of two compilation phases (\Cref{fig:arch}).
The first phase (projection) uses EPP to translate a choreography into an abstract representation of process programs given in SP, used as an intermediate representation (IR).
This phase is certified: it uses a Haskell program extracted from the Coq formalisation~\cite{CMP21b}.
The second phase (code generation) translates the abstract actions in the IR into an executable programming language.
Currently, we target Jolie.
This phase is not formally verified: since it is a homeomorphic transformation that follows the term structure of the IR in a straightforward fashion, its correctness is easy to establish directly by manual inspection.

Our development allows for writing and executing CC choreographies~\cite{CM20,CMP21b,CMP21,M23}.
Also, it confirms the informal claim that the formalisation of CC was made with flexibility in mind for this kind of applications~\cite{CMP22}.
In particular, our application to Jolie did not require any modification of the formalisation, but just an appropriate instantiation of its parameters and a simple interfacing of its extracted data types with the other components of our compilation pipeline in Haskell.

\paragraph*{Structure.}

\Cref{sec:arch} gives an overview of our compiler's architecture and describes its implementation.
\Cref{sec:comp} explains how we map the process language to Jolie in order to generate executable Jolie code consisting of multiple independent services.

\paragraph*{Related work.}

There are two other formalisations of choreographic programming.
Kalas~\cite{PGSN22} is a choreographic programming language formalised in the Hol4 proof assistant.
It comes with an end-to-end certified compiler that targets CakeML~\cite{MO12}, a formally-verified subset of the ML language.
By contrast, we take a technology-agnostic approach that allows for reusing our compiler infrastructure for different target languages.
Pirouette~\cite{HG22} is a functional choreographic programming language formalised in Coq, which similarly to~\cite{CMP21} can be instantiated with different languages for local computation.
However, it has not yet been used to implement a choreographic compiler that targets executable code.

\section{Choreographies in \hacc{}}
\label{sec:arch}

We represent choreographies and processes (terms of CC and SP, respectively) by Haskell data types that have been automatically extracted from the Coq formalisation.
However, a few considerations are necessary, given that Coq is a dependently-typed language, while all of the languages it can extract to are not.
As mentioned before, the formalisation is parametric over the types used to represent process identifiers, (recursion) variables, terms of the local computation language, etc.
Because this is done using the dependently-typed features of Coq, the extracted Haskell code has some peculiarities.

In particular, instead of using Haskell's type system and parametric polymorphism, the extracted code uses Haskell's \hc{Any} type to achieve genericity.
As a consequence, interfacing the extracted code requires using the \hc{unsafeCoerce} function.
To deal with the verbosity we provide a more ergonomic interface with a thin \textit{wrapper} (\Cref{fig:arch}) around the extracted code.
The wrapper hides the necessary coercions and models the terms using Haskell's parametric polymorphism.
This step is not formally verified, but the wrapper again follows the term structure of CC and SP, so checking it for correctness manually is not a problem.

\Cref{fig:iface-chor} shows our interface.
We fix the types of process identifiers (\hc{Pid}), variable names (\hc{Var}, \hc{RecVar}), selection labels (\hc{Label}) and annotations (\hc{Ann}) for simplicity.
The \hc{Label} type is a binary sum type with \hc{CLeft} and \hc{CRight} as its constructors, while others are wrappers around \hc{String}s, used as identifiers.

\begin{figure}
  \centering
  \begin{haskelllisting}
data Eta e = Com Pid e Pid Var | Sel Pid Pid Label

data Choreography e b
  = CEnd
  | Interaction (Eta e) Ann (Choreography e b)
  | CCond Pid b (Choreography e b) (Choreography e b)
  | CCall RecVar

newtype CDefSet e b = CDefSet [(RecVar, Choreography e b)]
newtype CProgram e b = CProgram (CDefSet e b, Choreography e b)
  \end{haskelllisting}
  \caption{Datatypes for choreographies in \hacc{}.}
  \label{fig:iface-chor}
\end{figure}

Choreographies have type \hc{Choreography e b}, parametric on the local computation and Boolean expression languages (\hc{e} and \hc{b}).
Processes can perform point-to-point interactions (\hc{Interaction})---value communications (\hc{Com}), where a process evaluates an expression and sends the result to another, or selections (\hc{Sel}), where a process selects how another process should behave by communicating a label.
Interactions include an annotation (\hc{Ann}), discussed below.
Conditionals (\hc{CCond}) are based on Boolean expressions, and \hc{CCall} invokes a named procedure.%
\footnote{CC includes runtime terms, needed for the semantics.
  Programmers should not write them explicitly, so we do not include them in our datatype.}
\hc{CProgram} is the type of choreographic programs, which includes definitions of named procedures (\hc{CDefSet}) as well as the main choreography.%
\footnote{In Coq, \cc{DefSet} also includes the set of processes involved in each procedure.
  In theory, this set might not be computable, as there may be infinitely many procedures.
  This cannot happen in hand-written choreographies, so our wrapper computes this set.}

Instead of working with our data types directly we define a convenient set of combinators for building choreographies.
We provide a \hc{prog} combinator that returns a \hc{CProgram} given a pair of recursive procedure definitions and the main choreography.
A choreography is given as a list of instructions (communications, selections of \hc{CLeft} or \hc{CRight}, conditionals and calls), all built using the corresponding combinators (\hc{com}, \hc{left}, \hc{right}, \hc{cond} and \hc{call}, respectively), which are strung together to produce a CC term in our representation.

When compiling to Jolie, the programmer can configure the generated code by annotating interactions using the \hc{ann} combinator.
These annotations will be used to override the default names of the Jolie operations that implement them.

\begin{figure}[b]
  \centering
  \begin{haskelllisting}[]
auth :: CProgram String String
auth = prog ([],
  [ann "authenticate" $\dollar$ com c "credentials" ip credentials,
   cond ip "check(credentials)"
     ([ann "authOk" $\dollar$ left ip s, ann "authOk" $\dollar$ left ip c,
       ann "acceptToken" $\dollar$ com s "makeToken" c token],
      [ann "authFail" $\dollar$ right ip s, ann "authFail" $\dollar$ right ip c])])
  where [ip, s, c] = pids ["Ip", "Server", "Client"]
        [credentials, token] = vars ["credentials", "token"]
  \end{haskelllisting}
  \caption{The distributed authentication choreography in CC.}
  \label{fig:ex-chor}
\end{figure}

\begin{example}[Distributed authentication]
  \label{ex:auth-chor}
  \Cref{fig:ex-chor} shows the distributed authentication choreography from \cite{CMP21b} encoded as a \hc{CProgram}.

  Here, \proc{Client} wishes to authenticate with \proc{Ip} (identity provider) in order to receive a token from \proc{Server}.
  \proc{Client} sends its credentials to \proc{Ip}, which checks them and communicates the result to both \proc{Client} and \proc{Server}.
  If the authentication was successful, \proc{Server} sends a token to \proc{Client}, otherwise the protocol terminates.

  The terms of the local computation and Boolean expression languages are simply strings, which are included as-is into the generated executable code.
  \eoe
\end{example}

\section{Compilation to Jolie}
\label{sec:comp}

The projected behaviour of a process has type \hc{Behaviour e b}, again parametric on the local languages (\Cref{fig:iface-proj}).
\begin{figure}
  \centering
  \begin{haskelllisting}
data Behaviour e b
  = BEnd
  | Send Pid e Ann (Behaviour e b)
  | Recv Pid Var Ann (Behaviour e b)
  | Choose Pid Label Ann (Behaviour e b)
  | Offer Pid (Maybe (Ann, Behaviour e b)) (Maybe (Ann, Behaviour e b))
  | BCond b (Behaviour e b) (Behaviour e b)
  | BCall (RecVar, Pid)

newtype BDefSet e b = BDefSet [((RecVar, Pid), Behaviour e b)]
newtype Network e b = Network [(Pid, Behaviour e b)]
newtype BProgram e b = BProgram (BDefSet e b, Network e b)

epp :: CProgram e b -> Maybe (BProgram e b)
  \end{haskelllisting}
  \caption{Datatypes for processes in \hacc{}.}
  \label{fig:iface-proj}
\end{figure}
\hc{Send} and \hc{Recv} correspond to the respective communication actions, while \hc{BCond} and \hc{BCall} are as in CC.
The remaining terms \hc{Choose} and \hc{Offer} are used to implement selections, where one process can \textit{choose} which of the \textit{offered} behaviours another process should execute.
Note that a process does not have to offer behaviours for both labels.

Processes are named by identifiers and grouped into a \hc{Network}, which is then paired with the projections of recursive procedures for each process (\hc{BDefSet}) to form a \hc{BProgram}.
Finally, \hc{epp} performs the projection of a \hc{CProgram} to a \hc{BProgram} using the extracted EPP as its foundation.
We use the \hc{Maybe} type to handle the case when a choreography is not projectable.

\begin{example}
  \label{ex:auth-proj}
  Projecting the distributed authentication choreography from \Cref{ex:auth-chor} gives us the process IR seen in \Cref{fig:ex-proj}.
  For conciseness, we only show the representative case of \proc{Client}.
  Note how the annotations from the choreography have been preserved and propagated to the projection.
  \begin{figure}
    \centering
    \begin{haskelllisting}[]
BProgram (BDefSet [], Network [...,
  (Pid "Client",
   Send (Pid "Ip") "credentials" (Ann "authenticate")
     (Offer (Pid "Ip")
       (Just (Ann "authOk",
              Recv (Pid "Server") (Var "token") (Ann "acceptToken") BEnd))
       (Just (Ann "authFail", BEnd))))])
    \end{haskelllisting}
    \caption{The projection of the choreography in \Cref{ex:auth-chor}.}
    \label{fig:ex-proj}
  \end{figure}
  \eoe
\end{example}

\begin{example}
  The process IR from \Cref{ex:auth-proj} can now be compiled down to executable code (\Cref{fig:ex-comp}).
  For each process, the backend generates a corresponding \jc{service} block in Jolie.
  We omit the necessary deployment configuration and show only the behaviour, whose structure follows that of the process IR.

  In this example, the annotations were used to specify the operation names exposed by the services (\jc{authOk} and \jc{acceptToken} for \proc{Client}, \jc{authenticate} for \proc{Ip}, etc.).
  In general, they can specify arbitrary metadata that allows the programmer to control and guide the backend's code generation.
  \begin{figure}
    \centering
    \begin{jolielisting}
service Client {
  ...
  main {
    authenticate@Ip( credentials )
    [ authOk() ] {
      acceptToken( token )
    }
    [ authFail() ] {
      nullProcess
    }
  }
}
    \end{jolielisting}
  \caption{The executable Jolie code generated for the choreography in \Cref{ex:auth-chor}.}
  \label{fig:ex-comp}
  \end{figure}
  \eoe
\end{example}

\section{Conclusion}

We implemented a toolchain for compiling choreographies to executable code in Jolie.
The complex step, computing the endpoint projection, is handled by certified code extracted from the Coq formalisation in~\cite{CMP21b}.
This certified code is then combined with uncertified wrappers whose correctness is easy to check by hand.
We illustrated the toolchain with a protocol for distributed authentication.

\bibliographystyle{splncs04}
\bibliography{main}

\end{document}